\documentclass[a4paper,12pt]{article}
\pdfoutput=1
\usepackage{cite}
\usepackage{amsmath}
\usepackage{amsfonts}
\usepackage{amssymb}
\usepackage{graphicx, rotating}
\usepackage{epsfig}
\usepackage{latexsym}
\usepackage{graphicx}
\usepackage{color}
\usepackage{amsmath,bm,amssymb}

\usepackage{color}
\usepackage[english]{babel}
\usepackage{hyperref}

\newcommand{\Slash}[1]{{\ooalign{\hfil#1\hfil\crcr\raise.167ex\hbox{/}}}}

\newcommand{\beq}{\begin{equation}}  \newcommand{\eeq}{\end{equation}}
\newcommand{\bef}{\begin{figure}}  \newcommand{\eef}{\end{figure}}
\newcommand{\bec}{\begin{center}}  \newcommand{\eec}{\end{center}}
  
\newcommand{\laq}[1]{\label{eq:#1}}  

\newcommand{\Eq}[1]{Eq.~(\ref{eq:#1})}

\def\({\left(}
\def\){\right)}

\def\O{\mathcal{O}}

\newcommand{\AND}{~{\rm and}~}
\newcommand{\EV}{ {\rm \, eV} }
\newcommand{\KEV}{ {\rm \, keV} }

\newcommand{\GEV}{ {\rm \, GeV} }

\def\d{\delta}

\def\f{\phi}
\def\g{\gamma}

\def\m{\mu}

\def\s{\sigma}

\def\D{\Delta}
\def\G{\Gamma}
\def\L{\Lambda}

\def\tl{\tilde}
\def\*{\dagger}

\begin{document}
\renewcommand\bibname{\Large References}

\begin{flushright}
TU-1154
\end{flushright}

\begin{center}

\vspace{1.5cm}

{\Large\bf Anisotropic cosmic optical background bound for \vspace{2.5mm}\\  decaying dark matter in light of the LORRI anomaly}\\
\vspace{1.5cm}

{\large \bf  Kazunori Nakayama$^{a,b}$} and {\large \bf Wen Yin$^{a}$}

\vspace{12pt}
\vspace{1.5cm}
{\em 
$^{a}${Department of Physics, Tohoku University, Sendai, Miyagi 980-8578, Japan} 
$^{b}${International Center for Quantum-field Measurement Systems for Studies of}\\
$^{ }${the Universe and Particles (QUP), KEK, 1-1 Oho, Tsukuba, Ibaraki 305-0801, Japan}\\
}

\vspace{1.5cm}
\abstract{
Recently anomalous flux in the cosmic optical background (COB) is reported by the New Horizon observations. The COB flux is
$16.37\pm1.47\, \rm nW m^{-2} sr^{-1}$, at the LORRI pivot wavelength of $0.608\,\rm \m m$, which is $\sim 4\s$ level above 
 the expected flux from the Hubble Space Telescope (HST) galaxy count. It would be great if this were a hint for the eV scale dark matter decaying into photons. 
In this paper, we point out that such a decaying dark matter model predicts a substantial amount of anisotropy in the COB flux, which is accurately measured by the HST. 
The data of the HST excludes the decay rate of the dominant cold dark matter larger than $10^{-24}$--$10^{-23}\,{\rm s}^{-1}$ in the mass range of $5$--$20$\,eV. As a result, the decaying cold dark matter explaining the COB excess is {strongly disfavored} by the anisotropy bound. 
We discuss some loopholes: e.g. warm/hot dark matter or two-step decay of the dark matter to explain the COB excess. 
}
\end{center}
\clearpage

\setcounter{page}{1}
\setcounter{footnote}{0}

\tableofcontents

\section{Introduction}

The origin of dark matter (DM) is one of the biggest mysteries in particle theory, astronomy, and cosmology. 
Many attempts are being made both from the theory side and experimental side. 
Recently the most precise measurement of the cosmic optical background (COB) was reported by 
the Long Range Reconnaissance Imager (LORRI) instrument on NASA's New Horizons mission~\cite{Zemcov:2017dwy, Lauer:2020qwk}.
The COB flux is measured to be $16.37\pm1.47 \rm nW m^{-2} sr^{-1}$, at the LORRI pivot wavelength of $0.608\rm \m m$
\cite{Lauer:2022fgc}. This is about $\sim 4\s$ level above the expected flux from the Hubble Space Telescope (HST) galaxy counts. It would be exciting if this is a hint of the DM~\cite{Lauer:2022fgc,Bernal:2022wsu}.

The sub keV (single-component) DM should be either a spin-zero or spin-unity bosonic particle but should not be the fermion due to the Tremaine-Gunn bound~\cite{Tremaine:1979we, Boyarsky:2008ju}. 
In the spin-zero case, the QCD axion that solves the strong CP problem~\cite{Peccei:1977hh,Peccei:1977ur,Weinberg:1977ma,Wilczek:1977pj} in the hadronic axion window may be a good candidate~\cite{Chang:1993gm, Moroi:1998qs}. Such a QCD axion as well as a more generic axion-like particle (ALP) can be produced non-thermally consistent with the cold DM paradigm~\cite{Mazumdar:2015pta,Co:2017mop,Moroi:2020has, Moroi:2020bkq, Nakayama:2021avl}.\footnote{
Thermally produced QCD axions produced by pion interactions are too abundant to be consistent with the cosmic microwave background observations unless the axion is lighter than $\sim 0.5$\,eV~\cite{Hannestad:2005df,Hannestad:2010yi,Archidiacono:2013cha,Archidiacono:2015mda,DiValentino:2015wba}. However, this bound is significantly relaxed for low enough reheating temperature~\cite{Grin:2007yg,Carenza:2021ebx}.
See also Refs.~\cite{DiLuzio:2021vjd,DEramo:2021psx,DEramo:2021lgb} for recent discussions about the theoretical estimation of thermal axion abundance.
}
Alternatively, the hypothesis that the inflaton and DM are unified by a single ALP predicts the
mass to be around eV~\cite{Daido:2017wwb, Daido:2017tbr} (see also Ref.\,\cite{IAXO:2019mpb}). 
In those cases, the sub-keV axions naturally decay into two photons with the photon couplings around $g_{\f \g\g}\sim 10^{-11}-10^{-10}\GEV. $ 
See Refs.~\cite{Jaeckel:2010ni,Ringwald:2012hr,Arias:2012az,Graham:2015ouw,Marsh:2015xka,Irastorza:2018dyq,DiLuzio:2020wdo} for reviews of the axion and ALPs.
In the ALP mass range of $\mathcal O(1$--$10)$ eV, the intensity of the cosmic optical background light was used to constrain the DM~\cite{Cadamuro:2011fd,Ringwald:2012hr}. In this paper, we may not regard it as a constraint since the excess in the optical background has been found~\cite{Lauer:2022fgc}. 
In such a situation, we may need to consider an independent constraint to check whether or how the decaying DM can explain the LORRI anomaly. 

In this paper we consider severe constraints on the DM scenario to explain the LORRI anomaly by using the COB anisotropy data~\cite{Kashlinsky:2018mnu}. Since the DM density spatially fluctuates in the universe, the photon flux from the DM decay not only contributes to the mean intensity but also to the anisotropy.
A similar analysis has been made in Refs.~\cite{Kohri:2017oqn, Kalashev:2018bra} for the ALP model to explain the cosmic infrared background (CIB) mean intensity excess observed by the CIBER experiment~\cite{Matsuura:2017lub}, whose wavelength is longer than the COB measured by LORRI.\footnote{
	See also Refs.~\cite{Gong:2015hke,Caputo:2020msf} for the ALP model to fit the CIB anisotropy data, rather than the CIB mean intensity. 
}
We apply the same idea to the DM model for the LORRI anomaly and derive constraints on such a scenario from the COB anisotropy measurements.

In this paper, we derive a robust bound from the anisotropic COB for simple cold DM models 
in the mass range of $\O(1$--$10)\EV$ (Fig.\ref{fig:2}), and show the exclusion limit for the ALP DM in the mass-photon coupling plane (Fig.\ref{fig:3}). 
 By taking account of the non-linear evolution of the density perturbation we found the bound from the data from HST~\cite{Mitchell-Wynne:2015rha} is so stringent that excludes all the parameter regions for the cold DM explanation of LORRI. 
We also discuss the possible loopholes and some more exotic DM models for explaining the LORRI anomaly.

\section{Isotropic and anisotropic extragalactic background light}   \label{sec:iso}

We introduce a particle $\chi$, which comprises a fraction $R$ $(\leq 1)$ of the total cold DM density, 
and it is assumed to have a decay mode into two particles including a photon $\g$:
\beq
\chi \to  \g +x
\eeq
where $x$ is a particle that may or may not be a photon. For simplicity we assume $x$ is massless.
We focus on the mass $m_\chi$ of $\chi$ in the range 
\beq
5\EV \lesssim m_\chi \lesssim 25\EV.
\eeq
As shown in the Figure later in this paper, the lower bound comes from the optical measurement of a galaxy~\cite{Regis:2020fhw}. 
We set the upper bound on the mass because of the severe constraint from the reionization history~\cite{Chen:2003gz,Zhang:2007zzh}. 
As noted in the Introduction, $\chi$ (and $x$) is unlikely to be a fermion due to the so-called Tremaine-Gunn bound~\cite{Tremaine:1979we, Boyarsky:2008ju}, $m_\chi \gtrsim 0.5\KEV$, which is derived from the upper bound on the phase space density in dwarf spheroidal galaxies if it is dominant and if the DM does not have intrinsic multiplicity. 
A nice candidate may be a spin-zero/two boson which can decay into a pair of photons or a photon plus an exotic vector boson or {
{a spin-one boson that can decay into a photon plus an exotic scalar boson. }
We assume the decay rate into photon is $\G$. 
Since it comprises a fraction $R$ of the cold DM, the averaged number density is expressed as $n_\chi=\rho_{\rm DM} R (1+z)^3 /m_\chi $, 
with $\rho_{\rm DM}$ being the measured  {present} energy density of DM, $z$ being the redshift. 
As we will see the parameters that are relevant in our analysis are as follows, as long as the $\chi$ lifetime is much longer than the age of Universe:
\begin{itemize}
\item $m_\chi $ which determines the wavelength of the resulting photon, $m_\chi/(2(1+z)).$
\item $\hat \G \equiv\G R  q_\g $. Here $q_\g= 1 \AND 2$ for $x$ is not $\g$  and is $\g$, respectively. 
\end{itemize}
The estimation of the photon flux does not depend on $R$, $q_\g$ and $\G$ independently, but depends on the combination $\hat \G$. 
{This means that our conclusions will also apply to the case that the dark matter is sub-dominant or/and decaying into a photon and a dark particle. However, it is not easily applied with $q_\g>2$ since the spectrum of the resulting photon will be quite different. }

As a concrete example model, we can consider an ALP as the dominant DM. 
In this case, 
\beq
 \G=\frac{ g_{\chi \g\g}^2}{64\pi} m_\chi^3 ~~~~~~\text{if DM is ALP},  \label{decayrate}
\eeq
with $g_{\chi \g\g}$ being the ALP photon coupling and $q_\g=2$.

\subsection{Formalism}

In this part we follow Refs.~\cite{Kohri:2017oqn} and \cite{Kalashev:2018bra} for calculating the extragalactic background light (EBL) from decaying particle.
As done in Ref.~\cite{Kalashev:2018bra}, we take the mean intensity of the flux detected at the energy $\omega$ with an observation bandwidth $\Delta\omega,$\footnote{
This averaging procedure is essential for estimating the anisotropy power spectrum in the case of line photon spectrum since otherwise the power spectrum would diverge at the observation frequency. There are several effects that smooth out the divergence such as the Doppler broadening due to the DM intrinsic velocity dispersion~\cite{Kohri:2017oqn}, but practically the effect of detector resolution at the observation energy band is much more important~\cite{Kalashev:2018bra}. Ref.~\cite{Kohri:2017oqn} overlooked this effect and overestimated the CIB anisotropy power by several orders of magnitude.
}
\begin{align}
\bar{I}(\omega,\Delta\omega)=\frac{1}{\Delta\omega}\int_{\Delta\omega} d\omega' \, \omega'^2 \int_z^{\infty}dz' W(z',\omega').
\end{align}
Here 
\begin{align}
\laq{Wz}
	W(z, \omega) \equiv \frac{1}{4\pi } \frac{\hat\G \rho_{\rm DM} }{H(z) m_\chi } \frac{dN_\gamma}{dE'},
\end{align}
where we have taken the speed of light to be unity. The Hubble parameter at the redshift $z$ is given by $H(z)=H_0 \sqrt{\Omega_\L+ \Omega_{m} (1+z)^3 +\Omega_{r} (1+z)^4}$ where $\Omega_\Lambda, \Omega_{m}$ and $\Omega_{r}$ denote the density parameter of the dark energy, total matter and radiation, respectively. 
The photon spectrum at the 2 body decay of $\chi$ has a delta-function shape\footnote{
	Note again that the photon multiplicity factor $q_\gamma$ (e.g. $q_\gamma=2$ for the ALP decay into two photons) is absorbed into the definition of $\hat \Gamma$.
}
\begin{align}
	\frac{dN_\gamma}{dE'} = \delta(E'-\omega_{\rm max}),
\end{align}
with $E'=(1+z)E$ and $\omega_{\rm max}=m_\chi/2$ in our massless approximation of the decay products. Since we are focusing on $\chi$ as (a part of) the DM, which has a lifetime longer than the age of the Universe, 
we made the {approximation $e^{-\G t} \to 1$ in the $\chi$ comoving number}. The exponential neglected in \Eq{Wz} is included in the numerical estimation.
The resulting isotropic COB energy flux is 
\beq
	\bar{I}(w,\Delta w)\simeq \omega^2\int dz W(z,\omega) = \frac{\omega }{4\pi } \frac{\hat\G \rho_{\rm DM}(0) }{H[z=\omega_{\rm max}/\omega-1 ] m_\chi } .   \label{flux}
\eeq
In order to explain the LORRI anomaly, we need $m_\chi=4$--$20\,\EV,$ with $\hat{\G}\sim 10^{-23}$--$10^{-22}\,{\rm s}^{-1}$~\cite{Bernal:2022wsu}.

The same setup also predicts the anisotropy of the photon flux since the DM density fluctuates in the universe~\cite{Ando:2005xg,Fornengo:2013rga}.
To discuss the anisotropy, let us expand the angular-dependent flux with spherical harmonics $Y_{\ell m}(\Omega)$,
\begin{align}
	\delta I (\omega,\D \omega, \Omega)  = I (\omega,\D \omega,\Omega) - \bar{I}(\omega,\Delta\omega) 
	= \sum_{\ell, m} a_{\ell m}(\omega, \Delta \omega)Y_{\ell m}(\Omega).
\end{align}
The relevant angular power spectrum is defined as \begin{equation}
C_\ell(\omega, \Delta\omega)=\langle|a_{\ell m}(\omega,\Delta\omega)|^2\rangle=\frac{1}{2\ell+1}\sum_{m=-\ell}^{+\ell}|a_{\ell m}(\omega, \Delta\omega)|^2
\end{equation}
{In terms of $W$ we obtain,}
\begin{align}
C_\ell(\omega,\Delta\omega)&=\frac{1}{\Delta\omega}\int_{\Delta\omega} d\omega_1 \, \omega_1^2 \int dz_1' W(z'_1,\omega_1)\nonumber\\
&\times\frac{1}{\Delta\omega}\int_{\Delta\omega} d\omega_2 \, \omega^2_2 \int dz_2' W(z'_2,\omega_2)\nonumber\\&
\times\frac{2}{\pi}\int dk k^2 P_\delta\left(k;r(z'_1),r(z'_2)\right) j_\ell(kr(z'_1))j_\ell(kr(z'_2))
\end{align}
with $r(z)=\int_0^z dz/H(z)$ being the comoving distance, and $j_\ell (kr(z))$ the spherical Bessel function. 
The power spectrum of the matter density fluctuation $\d$ is defined as \beq \left< \delta_{\vec k}(r) \delta_{\vec k'}(r') \right> = (2\pi)^3\delta^3(\vec k + \vec k') P_\delta(k; r,r').\eeq 
The power spectrum will be discussed in detail later. 
When the power spectrum varies slowly with $k$ the Limber approximation is applicable~\cite{Ando:2005xg,LoVerde:2008re}
\begin{align}
&\frac{2}{\pi}\int dk k^2 P_\delta(k;r(z'_1),r(z'_2))j_\ell(kr(z'_1))j_\ell(kr(z'_2))\nonumber\\
&\simeq\frac{1}{r(z_1')^2}  P_\delta\left(k=\frac{\ell}{r(z_1')};r(z'_1),r(z'_1)\right)\delta^{(1)}(r(z_1')-r(z_2'))  +\mathcal{O}(l^{-2}).
\end{align}
Defining $z^{\rm max}=\omega^{\rm max}/(\omega-\Delta\omega/2)-1$ and $z^{\rm min}=\omega^{\rm max}/(\omega+\Delta\omega/2)-1$ as the maximum and minimum redshift observed in the anisotropy measurement, {we have
\begin{align}
C_\ell(\omega,\Delta\omega)=&\int_{z^{\rm min}}^{z^{\rm max}}dz\left\{\frac{1}{4\pi}\frac{e^{-\Gamma t(z)}}{H(z)(1+z)^3}\omega_{\rm max}^2\hat \Gamma \frac{\rho_{\rm DM}}{m_\chi}\frac{1}{\Delta\omega}\right\}^2
\times \frac{H(z)}{r(z)^2} P_\delta (k=\frac{\ell}{r[z]}; r[z],r[z]) \ .
\label{Cell}
\end{align}}
Note that the integral depends on the observation frequency $\omega$. The observation bandwidth $\Delta\omega$ depends on the experimental setup.

The next task is to evaluate $ P_\delta\left(k,{r},r\right).$
To evaluate the power spectrum, we should take into account the non-linear structure formation effect.
We include the one and two-halo contributions~\cite{Cooray:2002dia}: 
\beq
P_\delta(k; r,r)=P^{\rm 1h}_\delta (k;r,r) +P^{\rm 2h}_\delta (k;r,r) 
\eeq
\begin{align}
\laq{1halo}
	P^{\rm 1h}_\delta (k;r,r) = \frac{1}{(\rho_m)^2}\int dM M^2\frac{dn(M,z)}{dM}|u_M(k)|^2,
\end{align}
and 
\begin{align}
	P^{\rm 2h}_\delta (k;r,r) =\left[\frac{1}{{\rho_m}} \int dM M\frac{dn(M,z)}{dM}u_M(k) b(M,z) \right]^2 P_\delta^{\rm (lin)}(k,z),
\end{align}
For small (large) distance scales the dominant one is the one-halo (two-halo) contribution. For comparison with the COB anisotropy data discussed later, the one-halo term is dominant for the most region of the relevant multipole moment $10^3\lesssim \ell\lesssim 10^6$. 
Since the estimation is complicated we list the various relevant functions and our strategy as follows. 

\begin{itemize}
\item  $dn/dM$ denotes the comoving number density of halo with the mass of $M$, 
\beq
\frac{dn}{dM}(M,z) = \frac{{\rho}_{m} }{M^2}\nu f(\nu) \frac{d \log \nu}{d \log M}
\eeq
where $\nu=\left[\delta_{\rm c}(z)/\sigma(M)\right]^2$, with the critical overdensity $\delta_{\rm c}(z)$ and $\sigma(M)$ is the variance of the linear density field in spheres containing a mean mass $M$,
{ \beq \sigma(M)^2=\frac{1}{2\pi^2}\int dk k^2 |\tl{W}(k R_M)|^2 P_\d^{\rm (lin)}(k)\eeq}
with $\tl W(x)=3(\sin x -x \cos x)/x^3$ being a top-hat window function. 
Also $R_M$ satisfies $(4\pi/3)R_M^3 \rho_m =M.$
We adopt the Sheth-Tormen form for the multiplicity function $f(\nu)$: \cite{Sheth:1999mn}
\beq
\nu f(\nu)=A\left(1+\frac1{\nu'^{p}}\right)\left(\frac{\nu'}{2\pi}\right)^\frac12 \, e^{-\nu'/2}
\eeq
where $\nu'=0.707\,\nu$, $p=0.3$ and $A=0.322$, which is fixed from $\int d\nu f(\nu)=1$.

\item $\rho_m$ is the averaged present (baryonic $+$ dark) matter energy density.
\item $P_\delta^{\rm (lin)}(k)$ is the linear matter density perturbation for which we use the output of the public code {\tt Class}~\cite{Blas:2011rf}  from the Planck best-fit cosmological parameters~\cite{Planck:2018vyg}. 
\item $b(M,z)$ is the linear halo bias~\cite{Cooray:2002dia}. 
\item $u_M(k)$ is the Fourier transform of the density profile of each halo~\cite{Cooray:2002dia}. For the Navarro-Frenk-White (NFW) density profile $\rho_{\rm dp}(r)=\rho_s r_s^3/r(r^2+r_s^2)$~\cite{Navarro:1995iw}, 
\begin{eqnarray}
 u_M(k) &=&  \frac{4\pi\rho_s r_s^3}{M}\, \Biggl\{ \sin(k r_s)\,
 \Bigl[ {\rm Si}([1+c_{\rm vir}]kr_s) - {\rm Si}(kr_s)\Bigr]
  -\frac{\sin(c_{\rm vir} kr_s)}{(1+c_{\rm vir})kr_s}  \nonumber \\
 && \qquad\qquad\qquad + \cos(kr_s)\,
               \Bigl[{\rm Ci}([1+c_{\rm vir}]kr_s)-{\rm Ci}(kr_s)\Bigr]  \Biggr\} \, , 
 \label{uknfw}
\end{eqnarray}
where the sine and cosine integrals are 
\begin{equation}
 {\rm Ci}(x) = -\int_x^\infty \frac{\cos t}{t}\,dt
 \quad{\rm and}\quad {\rm Si}(x) = \int_0^x \frac{\sin t}{t}\, dt \, ,
\end{equation}
\end{itemize}
For given halo mass $M$ and redshift $z$, the concentration parameter, $c_{\rm vir}$, and the parameters $r_s, \rho_s$ are obtained as follows.
First, let us define the Virial radius $R_{\rm vir}$ as
{\begin{align}
M&=\frac{4\pi}{3} R_{\rm vir}^3(M) \Delta_{\rm vir}(z) \bar\rho_m[z],
\end{align}}
{where~\cite{Bryan:1997dn}}
\begin{align}
\D_{\rm vir}(z)&=\frac{(18\pi^2+82y -39y^2) }{\hat\Omega_m(z)},~~~~\text{where}~~~~\hat\Omega_m(z)\equiv\frac{ \Omega_m (1+z)^3 H^2_0}{H(z)^2},
\end{align}
with $y=\hat\Omega_m(z)-1$. Using the Virial radius, the concentration parameter is defined as
\begin{align}
	c_{\rm vir}\equiv \frac{R_{\rm vir}}{r_{-2}}~~~~~~~\text{where}~~~~~~~\frac{d}{d r} (r^2 \rho_{\rm dp}(r))|_{r=r_{-2}}=0.
\end{align}
For the NFW profile, $r_s=r_{-2}$.
The $M,z$ dependence of the concentration function is model-dependent. For instance, in the power-law model~\cite{Neto:2007vq,Maccio:2008pcd, Cirelli:2010xx}
{\beq
 c_{\rm vir}(M,z)=6.5\,  \(H(z)/H_0\)^{-2/3} (M/M_*)^{-0.1}, M_*=3.4\times 10^{12} h^{-1}M_\odot.  \label{cvirz}
 \eeq}
By using this fitting function, we will obtain $r_s$. Finally, we obtain $\rho_s$ by the condition
\begin{align}
M=\int_0^{R_{\rm vir}} \rho_{\rm dp}(r) d^3r.
\end{align}

\subsection{Numerical result}  \label{sec:num}

So far, we see that a cold DM that explains the isotropic background flux necessarily induces an anisotropic one as well. 
Here we perform the numerical simulation to check how sizable the anisotropic contribution is. 
The $M$ integration is performed in the range $(10^{-6}-10^{17})M_\odot$ for numerical calculation with $M_\odot$ being the solar mass.
We also take $\Delta\omega=\omega$ as assumed in Ref.~\cite{Kalashev:2018bra}. {This actually is a conservative choice (see the last paragraph of this section).}
 
The resulting angular power spectrum of the COB is shown in Fig.\ref{fig:1} in $\ell$--$\ell^2 C_{\ell}/(2\pi) $ plane. 
We have taken $\lambda_{\rm obs}=0.85\,\m {\rm m}$\footnote{Strictly speaking $0.85\m $m may corresponds to CIB, but in this paper we also call it COB for simplification of the presentation.} in the left panel and $\lambda_{\rm obs}=0.606\,\m {\rm m}$ in the right panel. In each panel prediction from the decaying DM is shown for $m_\chi=10$\,eV and $15$\,eV with fixed $\hat{\G}=2\times10^{-23}$\,s$^{-1}$.
Also shown are the observed data points from the Hubble Space Telescope~\cite{Mitchell-Wynne:2015rha} with the error bars.

\begin{figure}[!t]
\begin{center}  
      \includegraphics[width=78mm]{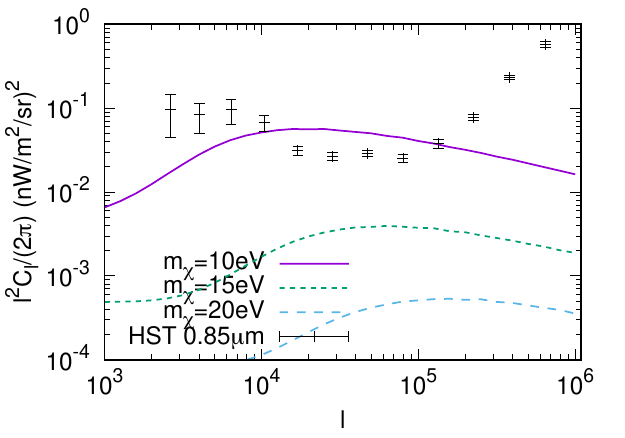}
      \includegraphics[width=78mm]{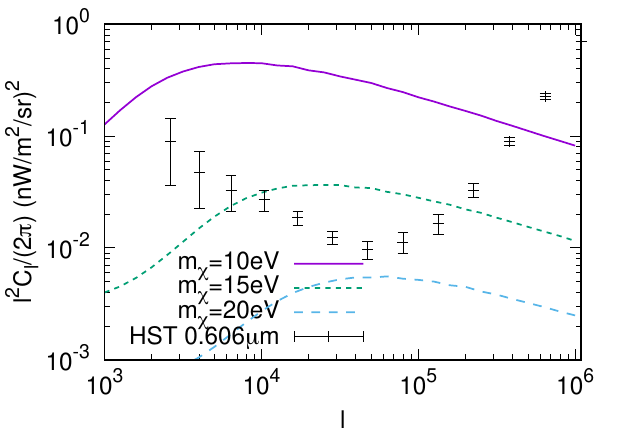}
      \end{center}
\caption{
Angular power spectrum of the COB anisotropy for the cold DM decaying into a mono-energetic photon. We have taken $\lambda_{\rm obs}=0.85\,\m {\rm m}$ in the left panel and $\lambda_{\rm obs}=0.606\,\m {\rm m}$ in the right panel. In each panel prediction from the decaying DM is shown for $m_\chi=10$\,eV, $15$\,eV and $20$\,eV with fixed $\hat{\G}=2\times10^{-23}$\,s$^{-1}$. Also shown are the data points from the HST observation.  
 } \label{fig:1}
\end{figure}

By requiring that the one-halo contribution in $C_{\ell}$ (see Eq.(\ref{Cell})) does not exceed the upper error bar of any of the data points for $\lambda_{\rm obs}=0.85\,\m$m and $\lambda_{\rm obs}=0.606\,\m$m, we derived the upper bound on $\hat \Gamma$ as shown in Fig.~\ref{fig:2}.
 Here we adopt the power-law model, which predicts $\mathcal O(0.1)$ smaller $C_\ell\ell^2$ from the previous analysis used in Fig.~\ref{fig:1}.  {The bound corresponds to the $95\%$CL exclusion limit by using a $\chi^2$ distribution, the degrees of freedoms of which are chosen as the number of center values of data points that are smaller than the model predictions at the corresponding $\hat\G$.\footnote{{We also checked that how we define statistics from the data is not very important in deriving the bound since the center values dominate over the error bars (see Fig.\ref{fig:1}). As we emphasized in the main text, the systematics are more important in deriving the bound, which we expect to strengthen the bound. }}} 
 One can see that even in this case, the region explaining the COB excess by the LORRI is, unfortunately {highly in tension with the COB anisotropy bound}. 
We also present the reionization bound~\cite{Cadamuro:2011fd,Ringwald:2012hr} and the indirect detection bound
from the observations of galaxy clusters, VIMOS Abell 2667 and 2390~\cite{Grin:2006aw}.\footnote{It is translated from the bound on $g_{\chi \gamma\gamma}$ taken from the webpage~https://cajohare.github.io/AxionLimits.}

In Fig.~\ref{fig:3} we translate the constraint on $\hat\Gamma$ from the COB anisotropy measurement in Fig.~\ref{fig:2} into the constraint on ALP-photon coupling $g_{\chi\gamma\gamma}$ for the ALP dominant DM, by using Eq.~(\ref{decayrate}). 
We also show the bound from the Horizontal branch star cooling for the photon coupling. We can see that the bound derived by us is more stringent than the cooling one. 

\begin{figure}[!t]
\begin{center}  
   \includegraphics[width=150mm]{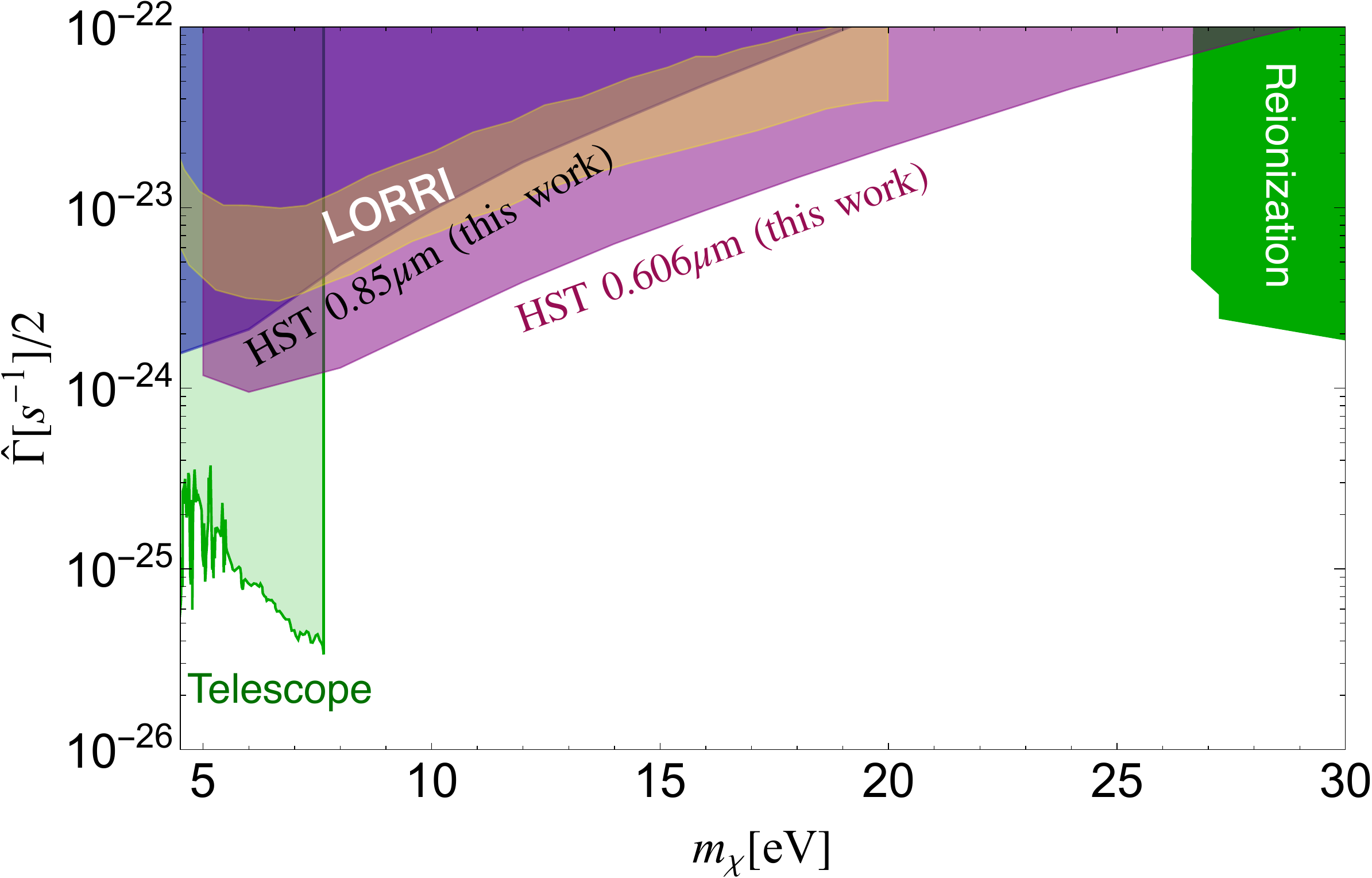}
      \end{center}
\caption{ The excluded region in ($m_\chi, \hat\G/2$) plane in a generic DM decaying into two particles, including a photon. 
The factor $2$ of the vertical line is taken so that it is the decay rate of the DM when $R=1,q_\g=2$.
The exclusion bound from the data of HST 0.85\,$\m$m and HST 0.606\,$\m$m are shown by darker blue and purple regions.   
Here $\hat \G=R q_\g \G$ as defined at the beginning of Sec.~\ref{sec:iso}. {The region that explains the LORRI anomaly is adopted from Ref.~\cite{Bernal:2022wsu}}
Also shown are the translated bounds from reionization and the optical telescope~\cite{Cadamuro:2011fd,Ringwald:2012hr,Grin:2006aw}.
 } \label{fig:2}
\end{figure}

\begin{figure}[!t]
\begin{center}  
      \includegraphics[width=150mm]{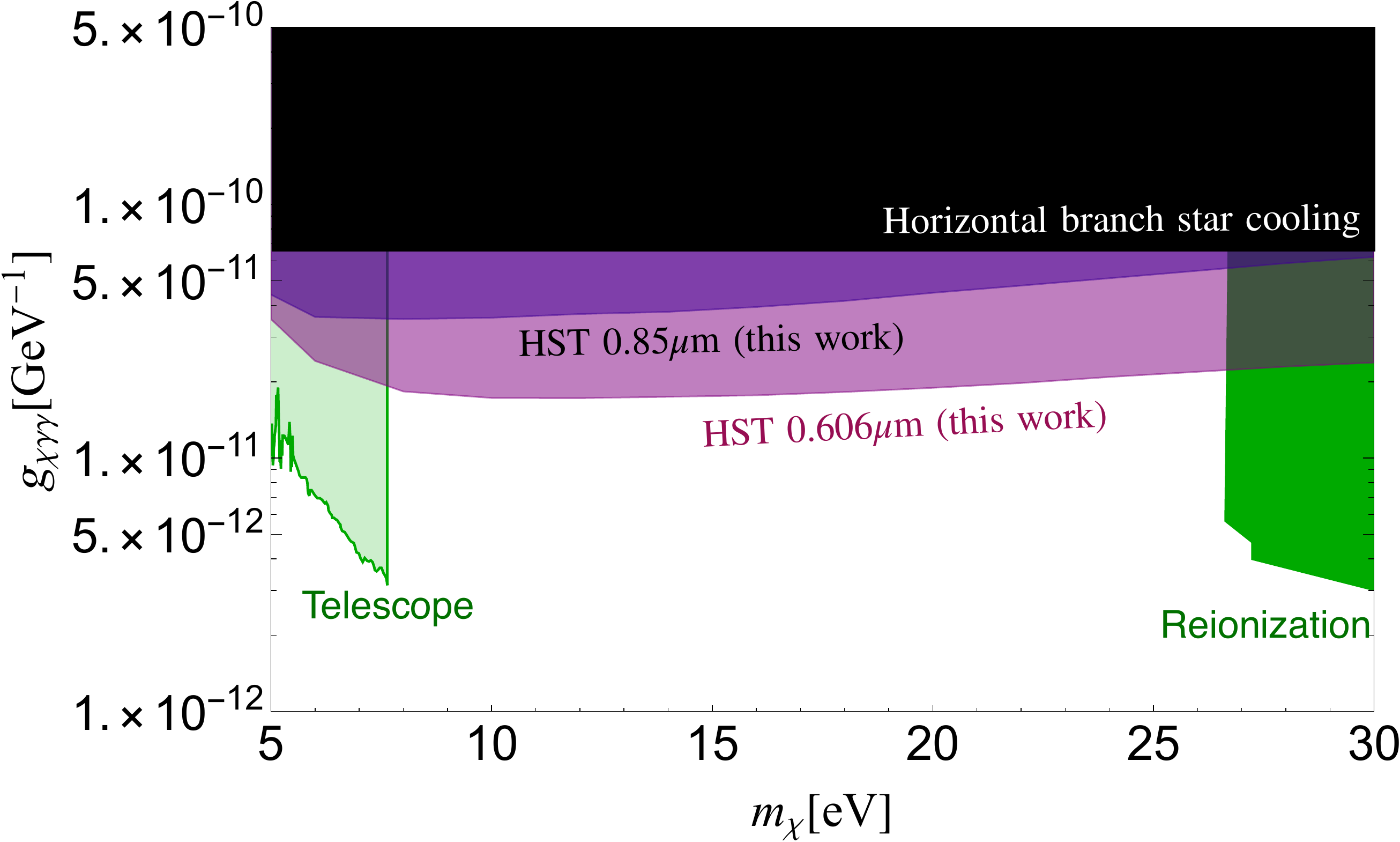}
      \end{center}
\caption{ Same figure as Fig.~\ref{fig:2} but we assume the ALP is the dominant cold DM, i.e. $R=1, q_\g=2$ and the constraint on $\hat\Gamma$ is translated into the ALP-photon coupling with Eq.~(\ref{decayrate}).
Also shown by the black region is the bound from the horizontal branch-star cooling~\cite{Raffelt:1985nk,Raffelt:1987yu,Raffelt:1996wa, Ayala:2014pea, Straniero:2015nvc, Giannotti:2015kwo}.} \label{fig:3}
\end{figure}

{Let us discuss uncertainties of the derived constraint. 
\begin{itemize}
\item We used the Sheth-Tormen fitting formula for the halo mass function, which should be compared with the result of N-body simulation. For parameter ranges we are interested in, the relevant redshift is $z\lesssim 3.9$ for $m_\chi \lesssim 20\,$eV for the observation frequency $\lambda= 0.606\,\mu$m. For such a redshift, the difference between the Sheth-Tormen fitting formula and the N-body simulation coincide within the accuracy of about factor two~\cite{Watson:2012mt}. Also we used the formula (\ref{cvirz}) for $c_{\rm vir}$. We checked that the formula given in Ref.~ \cite{Bullock:1999he} gives almost the same constraint.
\item As for the DM density profile, we employed the NFW one~\cite{Navarro:1995iw}. If we adopt other ones like the Burkert profile~\cite{Burkert:1995yz, Salucci:2000ps} (see also \cite{Salucci:2018hqu}), the resulting COB angular power spectrum may change only slightly at large $\ell$. We have also checked that the dominant one-halo contribution changes by at most $\O(1)$ factor at $\ell\lesssim 10^6$ by taking a lower cutoff of $M$ integration in \Eq{1halo} from $10^{-6}M_{\odot}$ to $10^9M_{\odot}$. Since the actual lower cutoff is expected to be much smaller than this value due to the smallness of the DM free streaming length and Compton wavelength, essentially, it does not lead to any uncertainty unless we assume a non-standard DM production scenario so that the free streaming length becomes extraordinarily long.
{For clarification, in Fig.~\ref{fig:M_P1h} we plot the relative contribution to the total one-halo power spectrum from logarithmic bins of the halo mass, $\left[d P^{\rm 1h}_\delta(k;M) / d\ln M\right]  / P_\delta^{\rm 1h}(k)$, which is defined by $P_\delta^{\rm 1h}(k) = \int d\ln M \left[dP^{\rm 1h}_\delta(k;M)/d\ln M \right]$, for $\lambda = 0.606\,\mu{\rm m}$ with $\ell=10^4$ (left) and $\ell=10^5$ (right). Note that the information of the wavelength $\lambda$ and the DM mass $m_\chi$ is relevant here since it determines the redshift $z$ at which the matter power spectrum is evaluated}, i.e., $\left[d P^{\rm 1h}_\delta(k;M) / d\ln M\right] = \left[d P^{\rm 1h}_\delta(k=\ell/r[z_{\rm dec}];M) / d\ln M\right]=
\rho_m^{-2} M^3\times {dn(M,z_{\rm dec})}/{dM}\times|u_M(k=\ell/r[z_{\rm dec}])|^2,
$ with $z_{\rm dec}=m_\chi \lambda /(4\pi)-1$.
\item For $m_\chi \gtrsim 15\EV$, the LORRI anomaly is explained by the low-frequency tail of the photon spectrum from the DM decay. Thus it predicts a much larger COB flux at a higher frequency. Such a case is highly disfavored by the TeV gamma-ray observation~\cite{HESS:2017vis,Fermi-LAT:2018lqt,Desai:2019pfl,MAGIC:2019ozu}. Although constraints from the TeV-gamma-ray observations are not shown in the Figure, one should note that the LORRI excess itself is slightly in tension with the TeV-gamma ray observations, and the tension becomes more and more serious for heavier axion. 
\item We also comment on the choice of the bandwidth $\Delta\omega$. It should be noticed that, according to Eq.\,\eqref{Cell}, $C_\ell \propto 1/\Delta \omega$ roughly, and hence the expected flux would become larger for narrower bandwidth for the case of line photon spectrum from decaying DM. Although we have chosen $\Delta\omega=\omega$ for numerical calculation, the actual bandwidth may be narrower by a factor $2$--$3$ for $\lambda=0.606\,\mu$m~\cite{Windhorst:2010ib}.
Thus the constraints that we have derived should be regarded as conservative ones in this respect.
In other words, the anisotropy measurement of the photon flux has the potential to greatly improve the constraint or the detection probability of decaying DM into a photon with a line spectrum.
\end{itemize}
}

\begin{figure}[!t]
\begin{center}  
      \includegraphics[width=78mm]{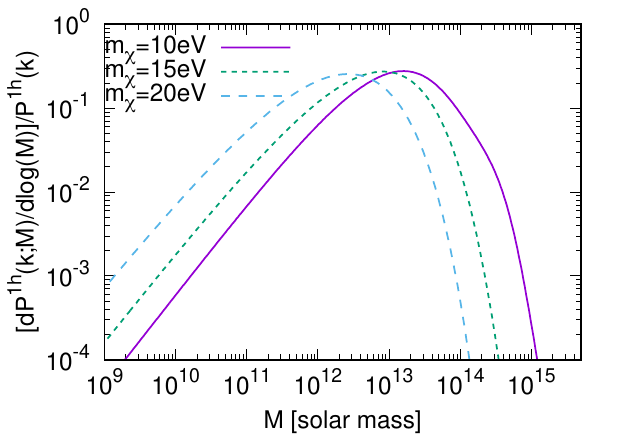}
      \includegraphics[width=78mm]{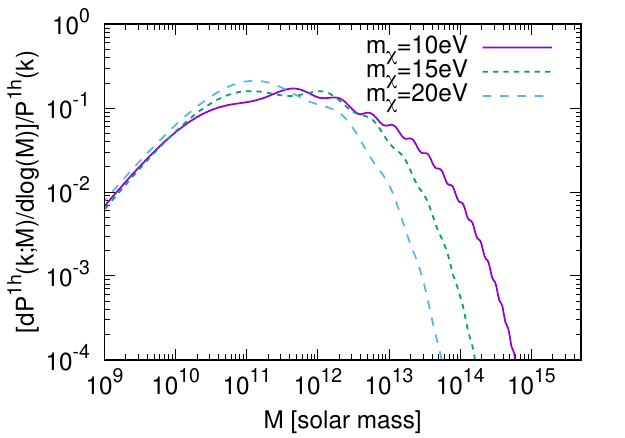}
      \end{center}
\caption{
	 The relative contribution to the total one-halo power spectrum from logarithmic bins of the halo mass, $\left[d P^{\rm 1h}_\delta(k;M) / d\ln M\right]  / P_\delta^{\rm 1h}(k)$, for $\lambda = 0.606\,\mu{\rm m}$ with $\ell=10^4$ (left) and $\ell=10^5$ (right). Note that the information of the wavelength $\lambda$ and the DM mass $m_\chi$ is relevant here since it determines the redshift $z$ at which the matter power spectrum is evaluated.
 } \label{fig:M_P1h}
\end{figure}

\section{Conclusions and discussion}

We derived upper bounds on the DM decay rate from the COB anisotropy in light of the excess in the COB mean intensity observed by LORRI. 
We found that the parameter region that explains the LORRI anomaly is {highly disfavored}
 by the COB anisotropy measurement. 
In other words, as long as $\chi$ plays the important role in the structure formation for the {parametrically} large scale, the derived bound is robust. 
Another remark is that the constraint we have derived should be regarded as a conservative one, as noted at the end of Sec.~\ref{sec:num}, due to the conservative choice of $\Delta\omega$. The constraint on $\hat\Gamma$ would become severer by a $\mathcal O(1)$ factor by precisely taking account of detector properties.

There are several loopholes to relax the COB anisotropy bound while keeping the mean intensity flux to explain the LORRI anomaly.
\begin{itemize}
\item \underline{Hot/warm $\chi$}: Let us suppose that $\chi$ has a relatively large velocity dispersion so that it is a hot/warm component rather than the cold one. If dominant, it is in tension with the structure formation and hence $R\ll 1$ is required. In particular, if $\chi$ has a larger velocity than the escape velocity of a DM halo, the magnitude of the anisotropy on small scales is reduced~\cite{Kalashev:2018bra}.\footnote{Given the 5~$\s$ Hubble tension, the bound on the hot DM may be different, c.f. the inflationary scenarios for explaining unusual primordial density perturbation required from various models for the Hubble tension~\cite{Takahashi:2021bti,DAmico:2021zdd} (See also \cite{Ye:2021nej, Vagnozzi:2021gjh} for the discussion to obtain $n_s \sim 1$ in order to alleviate the Hubble tension.)} 
As a simple scenario, we may consider the case that $\chi$ is thermalized in the very early Universe. Assuming that $\chi$ is a real scalar, we obtain the corresponding effective number of neutrino species $\Delta N_{\rm eff}\sim 0.027 (106.75/g^{\rm dec}_{\star})^{4/3}$, with $g^{\rm dec}_{\star}$ being the relativistic degrees of freedom at the decoupling of $\chi$ production. This scalar contributes to a fraction of the DM as a hot DM component, and the cosmological bound on such a hot relic reads $m_\chi \lesssim 10\EV$ for $g^{\rm dec}_{\star}=\O(100).$
 
\item \underline{Two step decay of $\chi$}: Cold $\chi$ decays to lighter mediator particles $\f_i$, and then the mediator decay into particles including a photon, 
\beq
\chi \to \sum_i \f_i\to  q_\g \times \g+ x.
\eeq
In particular, we are interested in the case where the mediator has a long enough lifetime so that the decay length of $\f_i$ is longer than the typical size of the DM halo and the typical distance between the halos. 
Then the resulting photons are almost isotropic due to the randomized second decay vertex positions (see Ref. c.f. \cite{Jaeckel:2021ert}).\footnote{A more exotic possibility may be the multi-component DM who are interacting with each other. In this case, the component that decays into particles, including a photon, may not follow the usual DM distribution, and hence the estimation of the anisotropy spectrum changes much.} In this sense, it may not even originate from the DM but from the decaying dark radiation from reheating~\cite{Jaeckel:2020oet, Jaeckel:2021gah} (see the case that the dark radiation is a decaying ALP \cite{Jaeckel:2021ert}. {The axion-photon conversion via cosmic magnetic field~\cite{Higaki:2013qka, Fairbairn:2013gsa, Conlon:2013txa, Tashiro:2013yea, Payez:2014xsa, Marsh:2017yvc, Reynolds:2019uqt} may also be a candidate to explain the LORRI anomaly, although it requires further study since the resulting photon has certain anisotropy due to the magnetic field distribution.}) In this case, one may confirm the evidence of the reheating by precisely determining the dark radiation spectrum~\cite{Jaeckel:2021gah} in the future line-intensity mapping experiments~\cite{Dore:2014cca,Hill:2008mv, Kovetz:2017agg, Bernal:2019jdo}. 
\end{itemize}
However, in the first case, we need to enhance $\chi$ coupling to a photon to compensate for {the small fraction, $R$,} and hence the bound from the stellar cooling becomes severer. 
On the other hand, it would be possible to modify the model to alleviate the stellar cooling bound~\cite{Kohri:2017oqn, Kalashev:2018bra, Jaeckel:2020oet, Masso:2005ym, Jaeckel:2006xm, Brax:2007ak}.
To summarize, the simple decaying DM into 2-particles, including a photon, is difficult to explain the LORRI anomaly, but there might be several loopholes.

\section*{Acknowledgement}
This work was supported by JSPS KAKENHI Grant Nos.  17H06359 (K.N.), 18K03609 (K.N.),  20H05851 (W.Y.), 21K20364 (W.Y.), 22K14029 (W.Y.), and 22H01215 (W.Y.).

\bibliographystyle{utphys}
\bibliography{ref}

\end{document}